\documentclass[11pt,a4paper]{article}
\usepackage{jheppub}
\usepackage[T1]{fontenc}
\usepackage{lmodern}
\usepackage{booktabs,mathtools,microtype,placeins}
\usepackage{flafter}
\renewcommand\afterProceedingsSpace{\vskip18pt}
\renewcommand\afterTitleSpace{\vskip18pt}
\renewcommand\afterRuleSpace{\vskip18pt}
\renewcommand\afterAuthorSpace{\vskip5pt}
\renewcommand\afterAffiliationSpace{\vskip3pt}
\renewcommand\afterEmailSpace{\vskip18pt}
\renewcommand\afterAbstractSpace{\vskip14pt}
\renewcommand\afterKeywordsSpace{\vskip14pt}
\newcommand{\dd}{\mathrm d}
\newcommand{\avg}[1]{\left\langle #1\right\rangle_z}
\newcommand{\eps}{\epsilon}
\newcommand{\cE}{\mathcal E}

\title{Striped phases of $\mathcal N=4$ SYM at finite density}
\author{Alex Buchel}
\affiliation{Department of Physics and Astronomy, University of Western Ontario,
London, Ontario N6A 5B7, Canada}
\emailAdd{abuchel@uwo.ca}
\abstract{We construct spatially modulated phases of strongly coupled
$\mathcal N=4$ supersymmetric Yang--Mills theory at finite $R$-charge density.
Compactification on a three-torus and unequal background charges isolate
an unstable charge-density mode. Its static threshold seeds a nonlinear
branch of striped black branes in the STU model. The striped states have
higher entropy than homogeneous states with the same energy and conserved
charges. They connect continuously to the homogeneous branch, with an
entropy excess that begins at fourth order in the modulation amplitude.
These solutions provide a candidate endpoint of the charge-transport
instability.}
\hypersetup{pdftitle={Striped phases of N=4 SYM at finite density},pdfauthor={Alex Buchel}}
\begin{document}
\maketitle
\clearpage

\section{Introduction}
\label{sec:introduction}
Thermodynamic instability can have a direct hydrodynamic manifestation.
For an uncharged plasma, the relation $c_s^2=\mathcal S/c_V$ connects the
speed of sound $c_s$ to the entropy density $\mathcal S$ and the specific
heat density $c_V$. Negative specific heat therefore turns a sound wave
into a growing energy-density and pressure fluctuation. In holography,
this identifies a dynamical instability of the dual black-brane
horizon~\cite{GM}. This connection provides a field-theory realization of
the correlated stability conjecture~\cite{GM1,GM2}, which proposes an
equivalence between dynamical and local thermodynamic stability for black
branes with noncompact translationally invariant directions. Local
thermodynamic stability requires the entropy to be concave as a function
of energy and conserved charges. When this condition fails, spatial
redistribution of these quantities can increase the entropy, motivating
an instability toward an inhomogeneous state. The sound-mode argument
makes this mechanism explicit for negative specific heat; at finite
charge density, unstable charge transport provides another possible
realization.

Such instabilities occur in many holographic theories. Top-down examples
include mass-deformed $\mathcal N=4$ plasma and cascading gauge
theory~\cite{N2thermo,N2critical,KS,Extended}. Charge-transport
instabilities also arise in conifold and M-theory
constructions~\cite{Baryonic,Mtheory,Membranes}. Bottom-up models provide
examples in which an unstable homogeneous plasma evolves into
inhomogeneous domains~\cite{Attems,Janik}. The connection between local
thermodynamics and long-wavelength dynamics is therefore a recurring
feature, while the equilibrium configurations that replace an unstable
state depend on the theory and its conserved quantities.

The Gladden--Ivo--Kovtun--Starinets (GIKS) instability concerns
$\mathcal N=4$ supersymmetric Yang--Mills theory with densities for the
three commuting $U(1)$ generators of its $R$ symmetry~\cite{GIKS,GIKSfollow}.
At equal charges, the homogeneous black brane becomes unstable to
charge-diffusion modes at sufficiently low temperature. There are two
independent unstable directions in charge space, and a finite band of
unstable spatial momenta. The instability thus destabilizes the
homogeneous and isotropic plasma without requiring unstable conformal
sound. At equal charges, the GIKS onset lies above the dimension-three
superconducting onset studied in \cite{Relevance}.\footnote{This
dimension-three channel arises from a bulk scalar carrying $R$-charge,
dual to a fermion bilinear in $\mathcal N=4$
SYM~\cite{Superstrings}. Its condensation spontaneously breaks the
corresponding $U(1)$ symmetry; this charged scalar is absent from the STU
truncation.} Other charged-scalar sectors have superconducting marginal
modes at higher temperatures~\cite{GentleWithers}; the present construction
concerns the charge-transport instability within STU.

Hydrodynamic clumping instabilities can lead to spatially inhomogeneous
equilibrium phases, as demonstrated by the nonlinear evolution of
holographic plasmas in the spinodal region~\cite{Attems,Janik}. The endpoint
of the GIKS charge-transport instability, however, remains unknown. Here
we construct its first spatially modulated candidate endpoint: a striped
phase with higher entropy than the homogeneous state at the same energy
and conserved charges.
To isolate a manageable sector, we place the theory on a rectangular
three-torus, choose two short transverse circles, and allow modulation
along the remaining circle. We also move away from the equal-charge point
so that the thermodynamic Hessian has only one negative eigenvalue. The
associated exchange-odd charge mode seeds a nonlinear striped branch.
Compactification on $S^3$, as in \cite{Sphere}, offers another way to
discretize the unstable modes, but would require a more involved nonlinear
construction. The torus preserves a simple Fourier description.

The striped solutions have larger entropy than homogeneous states at the
same energy, all three total charges and torus volume. This microcanonical
comparison is appropriate for future studies of isolated real-time
evolution, which conserves these quantities while allowing densities to
redistribute; see \cite{Janik} for a related holographic study of
spinodal evolution. The homogeneous ordered phase of \cite{Ordered,OrderedSTU}
was also compared microcanonically at fixed diagonal $R$-charge. That
comparison does not hold all three individual charges fixed; here we match
the energy, each charge, and the torus periods. We have not computed the dynamical
stability of the striped states; the additional modulation channels that
become available on larger tori are discussed in section~\ref{sec:discussion}.

The rest of the paper is organized as follows.
Section~\ref{sec:homogeneous} reviews homogeneous thermodynamics and the
unstable charge sector. Section~\ref{sec:stripes} develops the nonlinear
ansatz, its linear seed and thermodynamic observables, and outlines the
numerical method used.
The results are presented in section~\ref{sec:results}, followed by the
discussion. Appendix~\ref{app:stu} collects the STU action and homogeneous
solution, and appendix~\ref{app:numerics} gives the numerical implementation
and validation.

\section{Homogeneous thermodynamics and instability of charged \texorpdfstring{$\mathcal N=4$}{N=4} plasma at strong coupling}
\label{sec:homogeneous}
We consider $\mathcal N=4$ supersymmetric $SU(N)$ Yang--Mills theory at
large $N$ and strong 't Hooft coupling, with three conserved $U(1)$
$R$-charges.

The holographic description of this charged sector is provided by the
five-dimensional STU model~\cite{STU1,STU2,STU3}, a consistent truncation
of type IIB supergravity on $S^5$. Its bosonic fields are the metric,
three Abelian gauge fields and two neutral scalars. They are dual,
respectively, to the stress tensor, the three $U(1)$ $R$-symmetry currents
and two neutral dimension-two scalar bilinear operators of
$\mathcal N=4$ SYM. The model therefore captures the coupled energy,
charge and scalar fluctuations responsible for the instability considered
here. Consistency of the truncation ensures that its solutions uplift
to ten-dimensional supergravity. We collect the action, the homogeneous
black-brane solution and the holographic dictionary in
appendix~\ref{app:stu}, and review their thermodynamic consequences below.

Let $E,S,Q_I$ denote the energy, entropy and charges, and let
$V=L_xL_yL_z$ be the spatial volume. The homogeneous densities are
\begin{equation}
 \cE=\frac EV,\qquad \mathcal S=\frac SV,\qquad
 \rho_I=\frac{Q_I}{V},\qquad I=1,2,3.
 \label{eq:physicaldensities}
\end{equation}
We remove the common $N^2$ dependence by defining
\begin{equation}
 e=\frac{8\pi^2}{N^2}\cE,\qquad
 h=\frac{2\pi}{N^2}\mathcal S,\qquad
 q_I=\frac{8\pi^2}{N^2}\rho_I.
 \label{eq:densities}
\end{equation}
These are normalized densities; they retain their energy dimensions.
In particular, $h$ is an entropy variable and is distinct from the
gravitational scale $s$ introduced below. The first law at fixed torus
periods is
\begin{equation}
 \dd e=4\pi T\,\dd h+\sum_I\mu_I\,\dd q_I.
 \label{eq:firstlaw}
\end{equation}

The homogeneous STU solution is specified by $s>0$ and three nonnegative
parameters $\kappa_I$. Define $P=\prod_I(1+\kappa_I)$. Its thermodynamics is
\begin{align}
 e&=3s^4P,&h&=s^3\sqrt P,&q_I&=s^3\sqrt{2\kappa_IP},\\
 T&=\frac{s(2+\sum_I \kappa_I-\prod_I \kappa_I)}{2\pi\sqrt P},&
 \mu_I&=\frac{s\sqrt{2\kappa_IP}}{1+\kappa_I}.&&
 \label{eq:homthermo}
\end{align}
We restrict to nonextremal outer horizons, for which
$2+\sum_I\kappa_I-\prod_I\kappa_I>0$.
The homogeneous pressure is $\cE/3$. Eliminating the gravitational
parameters gives the equation of state
\begin{equation}
 e(h,q_I)=3h^{4/3}\prod_{I=1}^3
 \left(1+\frac{q_I^2}{2h^2}\right)^{1/3},\qquad
 \kappa_I=\frac{q_I^2}{2h^2}.
 \label{eq:eos}
\end{equation}
Differentiating this expression reproduces
$\partial_h e=4\pi T$ and $\partial_{q_I}e=\mu_I$.

Local thermodynamic stability requires convexity of $e$ in
$Y=(h,q_1,q_2,q_3)$, or positivity of
\begin{equation}
 \mathsf H_{AB}=\frac{\partial^2e}{\partial Y_A\partial Y_B}.
 \label{eq:hessian}
\end{equation}
For example, write $\ell_A=\partial_A\log e$ and
$a=(4-2\sum_I \kappa_I/(1+\kappa_I))/3$. Then
\begin{align}
 \ell_h&=\frac ah,&\ell_I&=\frac{q_I}{3h^2(1+\kappa_I)},\\
 \partial_h\ell_h&=\frac{-a+\frac43\sum_I \kappa_I/(1+\kappa_I)^2}{h^2},&
 \partial_{q_I}\ell_h&=-\frac{2q_I}{3h^3(1+\kappa_I)^2},\\
 \partial_{q_J}\ell_I&=\delta_{IJ}\frac{1-\kappa_I}{3h^2(1+\kappa_I)^2},&
 \mathsf H_{AB}&=e(\ell_A\ell_B+\partial_A\ell_B).
 \label{eq:hessianexplicit}
\end{align}
These formulas permit a direct separation of the unstable sectors.

We choose $\kappa_1=\kappa_2$, keeping $\kappa_3$ independent. The exchange $1\leftrightarrow2$
is then an exact symmetry of the background. The direction
$v_-=(0,1,-1,0)/\sqrt2$ is a Hessian eigenvector with eigenvalue
\begin{equation}
 \lambda_-^{\rm th}=\frac{e}{3h^2}\frac{1-\kappa_1}{(1+\kappa_1)^2}.
 \label{eq:oddhessian}
\end{equation}
Thus $\kappa_1>1$ destabilizes the charge-density difference $\rho_1-\rho_2$.
The remaining three eigenvalues can stay positive when $\kappa_3\ne \kappa_1$;
our results use such a point. At the fully symmetric point the two
independent charge-difference directions are degenerate, giving a larger
unstable sector.

A negative diffusion coefficient gives a growing mode with
$\omega=-iD k^2+\cdots$. The full gravitational fluctuation problem
determines where this growth stops at finite momentum. We denote its
static upper threshold by $k_c$. The thermodynamic Hessian diagnoses the
long-wavelength instability; it does not determine $k_c$.
The follow-up analysis of \cite{GIKSfollow} includes the asymmetric
case $(\kappa_1,\kappa_2,\kappa_3)=(\kappa,\kappa,0)$, with one unstable charge-difference
mode for $\kappa>1$ and a finite unstable band.

On a circle the nonzero momenta are $k_n=2\pi n/L_z$. The threshold of
the first longitudinal mode is therefore $L_{z,c}=2\pi/k_c$ for a specified
homogeneous state. Taking $L_x,L_y$ sufficiently short removes transverse
momenta from the same band. In what follows, $s_c,\kappa_{1,c},\kappa_{3,c}$ label the
onset of the striped branch, where it bifurcates from the homogeneous
solution; $k_c$ denotes the critical spatial wave number.

\section{Constructing striped states}
\label{sec:stripes}
We construct striped states by allowing the bulk fields to depend on the
radial coordinate and one boundary spatial direction.
Section~\ref{sec:ansatz} introduces the nonlinear ansatz and its boundary
and horizon conditions, while section~\ref{sec:symmetries} specifies the
discrete symmetries and Fourier expansions. Section~\ref{sec:linear}
derives the linearized mode that seeds the striped branch. We then obtain
the nonlinear thermodynamic observables and explain the microcanonical
entropy comparison in section~\ref{sec:thermo}, before outlining the
numerical method in section~\ref{sec:method}.

\subsection{Nonlinear ansatz and regularity}
\label{sec:ansatz}
We seek states invariant under translations in $x,y$, with equal metric
components in those directions, and periodic in $z$. The computational
domain is
\begin{equation}
 0\le r\le1,\qquad z\sim z+L_z,\qquad
 \theta=k_cz\sim\theta+2\pi,\qquad L_z=\frac{2\pi}{k_c}.
 \label{eq:domain}
\end{equation}
The conformal boundary is $r=0$ and the connected horizon is $r=1$.
The angular variable $\theta$ obeys $\partial_z=k_c\partial_\theta$.

We restrict to the consistent static, purely electric sector
\begin{equation}
 A^I=\Phi_I(r,z)\,\dd t,\qquad
 \phi=\phi(r,z),\qquad\chi=\chi(r,z).
 \label{eq:electric}
\end{equation}
Staticity means that the timelike Killing field is hypersurface orthogonal.
For a metric written with a time one-form
$\dd t+w_r\dd r+w_z\dd z$, this gives
$\partial_rw_z-\partial_zw_r=0$. If the accumulated time shift around
the circle vanishes, $\int_0^{L_z}w_z\,\dd z=0$, a single-valued
redefinition of $t$ removes both shifts. The mixed
time--space Einstein equations and the spatial Maxwell equations then
vanish identically. We also set to zero the spatial field strengths and
the Wilson lines, which describe constant spatial gauge potentials with
nonzero integrals around torus cycles. These choices are consistent with
the retained equations. The Chern--Simons sources
vanish because $F^I\wedge F^J=0$.

The metric is parametrized as
\begin{equation}
 \dd s_5^2=-\bar N^2e^{2U_t}\dd t^2+
 \bar G_s e^{2U_p}(\dd x^2+\dd y^2)
 +\bar G_s e^{2U_z}(\dd z+W\dd r)^2
 +\bar G_r e^{2U_r}\dd r^2.
 \label{eq:metric}
\end{equation}
All five deformation functions depend on $(r,z)$. The reference functions
are those of a homogeneous STU brane:
\begin{align}
 H&=(1+\kappa_1r)^2(1+\kappa_3r),&P&=(1+\kappa_1)^2(1+\kappa_3),&f&=H-r^2P,\\
 \bar N^2&=\frac{s^2f}{rH^{2/3}},&
 \bar G_s&=\frac{s^2H^{1/3}}r,&
 \bar G_r&=\frac{H^{1/3}}{4r^2f}.
 \label{eq:reference}
\end{align}
The reference scalar and electric potentials are given in
appendix~\ref{app:stu}, with $\kappa_2=\kappa_1$ and $\bar\chi=0$.

We build the leading boundary and horizon conditions into auxiliary
functions, each depending on $(r,z)$:
\begin{align}
 U_r&=r^2\mathcal R,&U_t&=r^2[\mathcal R+(1-r)\mathcal T],&
 U_p&=r^2\mathcal P,\\
 U_z&=r^2\mathcal Z,&W&=r^2\mathcal W,&&\\
 \phi&=\bar\phi+r\mathcal F,&\chi&=r\mathcal C,&&\\
 \Phi_+&=\bar\Phi_1+r(1-r)\mathcal A,&
 \Phi_-&=r(1-r)\mathcal M,&
 \Phi_3&=\bar\Phi_3+r(1-r)\mathcal B,
 \label{eq:auxiliaries}
\end{align}
where $\Phi_\pm=(\Phi_1\pm\Phi_2)/2$. The auxiliary functions are bounded
at the boundary, with the falloffs described below, and analytic in
$y=1-r$ at the horizon. The scalar logarithmic sources are absent.
Their normalizable coefficients, as well as the boundary electric slopes,
can vary with $z$. The boundary metric is flat: its representative follows
from the leading defining function $u=\sqrt r/s$.

At the horizon, write $y=\varrho^2$ and
$G=2+2\kappa_1+\kappa_3-\kappa_1^2\kappa_3$. The Euclidean radial--time metric is
\begin{equation}
 \dd s^2_{\tau\varrho}\simeq
 \frac{s^2G}{P^{2/3}}e^{2U_t(1,z)}\varrho^2\dd\tau^2
 +\frac{P^{1/3}}G e^{2U_r(1,z)}\dd\varrho^2.
 \label{eq:conical}
\end{equation}
Since $U_t(1,z)=U_r(1,z)$, one Euclidean time period removes the conical
singularity everywhere. With $\partial_t$ normalized by boundary time,
the surface gravity and temperature are
\begin{equation}
 \kappa_H=\frac{sG}{\sqrt P}=2\pi T,\qquad \partial_z\kappa_H=0.
 \label{eq:temperature}
\end{equation}
The horizon area density can be modulated while its surface gravity is
constant, in agreement with the zeroth law of black-hole mechanics.
Its value can change between members of the branch.
The electric potentials vanish as $\varrho^2$, so their Euclidean
one-forms are regular at the contractible time circle.

\subsection{Discrete symmetries and Fourier fields}
\label{sec:symmetries}
The internal exchange $\mathsf P_{12}$ sends
$\chi\to-\chi$ and $\Phi_-\to-\Phi_-$, leaving the metric, $\phi$,
$\Phi_+$ and $\Phi_3$ unchanged. The linear instability belongs to this
odd sector. We impose invariance under the \emph{combined} operation
\begin{equation}
 \mathsf P_{12}:1\leftrightarrow2,\qquad z\longmapsto z+L_z/2.
 \label{eq:combined}
\end{equation}
A half-period translation reverses the odd fields once more, restoring
the configuration. Consequently the even fields have period $L_z/2$,
while the odd fields change sign under this translation.

We restrict to reflection-symmetric stripes and choose their reflection
center at $z=0$. Under $z\to-z$, scalar functions and
electric potentials are even under this reflection, whereas the metric
coefficient $W$ is odd because it multiplies the $\dd r\,\dd z$ term.
It is even under the half-period translation. The explicit expansions are
\begin{align}
 (\mathcal C,\mathcal M)(r,z)
 &=\sum_{j=0}^\infty(C_{2j+1},M_{2j+1})(r)
          \cos[(2j+1)k_cz],\\
 \mathcal V(r,z)&=\sum_{j=0}^\infty V_{2j}(r)\cos(2jk_cz),
 \quad\mathcal V\in\{\mathcal T,\mathcal R,\mathcal P,
              \mathcal Z,\mathcal F,\mathcal A,\mathcal B\},\\
 \mathcal W(r,z)&=\sum_{j=1}^\infty W_{2j}(r)\sin(2jk_cz).
 \label{eq:fourier}
\end{align}
Thus $W$ contains even sine harmonics, with no zero mode. The odd
perturbations begin at order $\eps$, and backreaction on the remaining
fields begins at order $\eps^2$. We define the amplitude by
\begin{equation}
 C_1(1)=\eps.
 \label{eq:amplitude}
\end{equation}
Changing its sign amounts to a half-period translation or internal exchange.
Spatially averaged thermodynamic quantities are therefore even in $\eps$.

\subsection{The linear seed}
\label{sec:linear}
On a homogeneous background with $\kappa_1=\kappa_2$, the odd sector decouples from
the metric and the even matter fields at first order. Define dimensionless
radial profiles by
\begin{equation}
 \chi=\eps\psi(r)\cos(kz)+O(\eps^3),\qquad
 \Phi_-=s\eps a(r)\cos(kz)+O(\eps^3).
 \label{eq:linearansatz}
\end{equation}
At onset $k=k_c$ and $\psi(1)=1$, in agreement with
\eqref{eq:amplitude}. Write $H_1=1+\kappa_1r$, $B=\sqrt{\kappa_1P}$ and
$K=k/s$. Primes in this subsection denote $r$ derivatives. The two static
equations are
\begin{align}
 0={}&a''+\frac{2\kappa_1}{H_1}a'
       -\frac{2B}{H_1^2}\psi'-\frac{K^2}{4rf}a,
       \label{eq:lineara}\\
 0={}&\psi''+\left(\frac{f'}f-\frac1r\right)\psi'
       -\frac{2Br}{f}a'
       +\left(\frac{H_1}{fr^2}
        +\frac{2\kappa_1Pr}{fH_1^2}-\frac{K^2}{4rf}\right)\psi.
       \label{eq:linearpsi}
\end{align}
The coefficient of the non-normalizable scalar behavior $r\log r$ is
the source for the dimension-two boundary operator. When this source
vanishes, the normalizable coefficient multiplying $r$ determines the
operator's expectation value. For the electric perturbation, the constant
boundary value of $a$ changes the chemical-potential difference, while
its radial slope determines the charge-density difference. Setting both
sources to zero gives
\begin{align}
 a&=a_b r+\left[B\psi_b+
                  \left(\frac{K^2}{8}-\kappa_1\right)a_b\right]r^2+O(r^3),\\
 \psi&=\psi_b r+\left(\frac{K^2}{4}-\kappa_1\right)\psi_b r^2+O(r^3).
 \label{eq:linearbdy}
\end{align}
The coefficients $a_b,\psi_b$ are independent normalizable data. With
$y=1-r$, a regular horizon expansion is
\begin{align}
 a&=a_h y+\left[\frac{\kappa_1a_h}{1+\kappa_1}
       -\frac{B\psi_{h,1}}{(1+\kappa_1)^2}
       +\frac{K^2a_h}{8G}\right]y^2+O(y^3),\\
 \psi&=\psi_h+\psi_{h,1}y+O(y^2),\\
 \psi_{h,1}&=\frac{[K^2-4(1+3\kappa_1+2\kappa_1\kappa_3)]\psi_h-8Ba_h}{4G}.
 \label{eq:linearhor}
\end{align}
Regularity leaves $a_h,\psi_h$ free; the overall normalization fixes
$\psi_h=1$. Higher coefficients at either endpoint follow recursively.

We integrate linearly independent solutions satisfying the source-free
boundary conditions from the boundary, and solutions satisfying regularity
at the horizon from the horizon. At an intermediate radial point, we match
the fields $a,\psi$ and their first derivatives. This gives a homogeneous
linear matching problem. A nontrivial null vector
selects $k_c$, together with the relative coefficients of the two fields.
This is a static eigenvalue calculation. In the symmetric limit and the
$\kappa_3=0$ limit, its threshold should reproduce the corresponding
zero-frequency GIKS fluctuation problem after matching units. The latter
benchmark is supplied explicitly in section 5.4 of \cite{GIKSfollow};
our unequal, nonzero third-charge background is a different point in the
thermodynamic family.

\subsection{Nonlinear thermodynamics and microcanonical matching}
\label{sec:thermo}
The natural comparison for an isolated evolution is microcanonical.
With fixed torus geometry and no time-dependent driving, the total energy
and each total charge are conserved. Their local densities can redistribute,
and the temperature and chemical potentials of the final equilibrium need
not equal their initial values. A candidate endpoint should therefore be
compared by its entropy at the same $E,Q_I,V$.

For periodic functions define
$\avg{F}=L_z^{-1}\int_0^{L_z}F(z)\,\dd z$.
The temperature follows from \eqref{eq:temperature}. Integrating the horizon
area element gives
\begin{equation}
 h=s^3\sqrt P\avg{e^{2U_p(1,z)+U_z(1,z)}}.
 \label{eq:entropy}
\end{equation}
The chemical potential is the boundary-to-horizon electric potential
difference. In the regular gauge \eqref{eq:auxiliaries},
\begin{equation}
 \mu_1=\mu_2=\frac{s\sqrt{2\kappa_1P}}{1+\kappa_1},\qquad
 \mu_3=\frac{s\sqrt{2\kappa_3P}}{1+\kappa_3}.
 \label{eq:nonlinearmu}
\end{equation}
These potentials are constant in space but vary with the solved global
parameters $s,\kappa_1,\kappa_3$. Fixing their reference-field form does not fix their
values along the branch.

Let $\rho_I(z)$ be the local physical charge density and put
$\widetilde\rho_I=16\pi G_5\rho_I$. The Maxwell normalization in
appendix~\ref{app:stu} gives the boundary flux
\begin{equation}
 \widetilde\rho_I(z)=\lim_{r\to0}
 \frac12\sqrt{-g}\,Z_IF^{I\,rt}
 =-s^2\partial_r\Phi_I(0,z),\qquad Z_I=(X^I)^{-2}.
 \label{eq:flux}
\end{equation}
Here $r$ increases towards the horizon. Substitution of the potentials yields
\begin{align}
 \widetilde\rho_1&=s^3\sqrt{2\kappa_1P}-s^2(\mathcal A_0+\mathcal M_0),\\
 \widetilde\rho_2&=s^3\sqrt{2\kappa_1P}-s^2(\mathcal A_0-\mathcal M_0),\\
 \widetilde\rho_3&=s^3\sqrt{2\kappa_3P}-s^2\mathcal B_0,
 \qquad q_I=\avg{\widetilde\rho_I}.
 \label{eq:localcharge}
\end{align}
The subscript $0$ on an auxiliary function means its boundary value at
$r=0$, still a function of $z$, and does not denote a Fourier zero mode.
The odd field averages to zero, so $q_1=q_2$ although their local densities
differ. Maxwell's equation implies
$\partial_r\avg{\frac12\sqrt{-g}Z_IF^{I\,rt}}=0$.

For the energy and stresses, define the boundary coefficients
\begin{equation}
 U_a=r^2u_{a,2}(z)+o(r^2),\quad a=t,r,p,z,\qquad
 \phi=r\phi_1(z)+o(r),\quad\chi=r\chi_1(z)+o(r).
 \label{eq:uvcoeff}
\end{equation}
Thus $\phi_1=\bar\phi_1+\mathcal F_0$, $\chi_1=\mathcal C_0$ and
$\bar\phi_1=\sqrt{2/3}(\kappa_1-\kappa_3)$. In Fefferman--Graham coordinates $u$,
\begin{equation}
 \dd s^2=\frac{\dd u^2}{u^2}
 +\frac{\eta_{ij}+u^4g_{(4)ij}+\cdots}{u^2}\dd x^i\dd x^j,
 \qquad (\phi,\chi)=u^2(\phi_{(2)},\chi_{(2)})+\cdots.
\end{equation}
For the flat boundary and vanishing scalar sources, holographic
renormalization gives~\cite{BFS}
\begin{equation}
 16\pi G_5\langle T_{ij}\rangle
 =4g_{(4)ij}+\tfrac13(\phi_{(2)}^2+\chi_{(2)}^2)\eta_{ij}.
 \label{eq:stress}
\end{equation}
The scalar contribution is required by the renormalized stress tensor.
Relative to the homogeneous reference, the leading radial correction is
$r=\bar r(u)[1-u_{r,2}\bar r(u)^2/2+\cdots]$, with
$\bar r(u)=s^2u^2+\cdots$. Consequently
$\delta g_{(4)tt}=-s^4(2u_{t,2}+u_{r,2}/2)$ and
$(\phi_{(2)},\chi_{(2)})=s^2(\phi_1,\chi_1)$.
Writing $D_\phi=\phi_1^2+\chi_1^2-\bar\phi_1^2$, we obtain
\begin{align}
 \widetilde\cE(z)&\equiv16\pi G_5\langle T_{tt}\rangle
       =s^4[3P-8u_{t,2}-2u_{r,2}-D_\phi/3],\\
 \widetilde p_\perp(z)&\equiv16\pi G_5\langle T_{xx}\rangle
       =s^4[P+8u_{p,2}+2u_{r,2}+D_\phi/3],\\
 \widetilde p_z(z)&\equiv16\pi G_5\langle T_{zz}\rangle
       =s^4[P+8u_{z,2}+2u_{r,2}+D_\phi/3],\qquad
 e=\avg{\widetilde\cE}.
 \label{eq:localstress}
\end{align}
The scalar squares are formed before averaging. The flat, source-free
Ward identities are $-\widetilde\cE+2\widetilde p_\perp+\widetilde p_z=0$
and $\partial_z\widetilde p_z=0$. They allow a modulated energy density
and anisotropic stresses.

For each striped solution we reconstruct its homogeneous comparator from
\eqref{eq:eos} at the same $e,q_I$. Its entropy satisfies
\begin{equation}
 \prod_{I=1}^3\left(h_{\rm hom}^2+\frac{q_I^2}{2}\right)
 -\frac{e^3}{27}h_{\rm hom}^2=0.
 \label{eq:cubic}
\end{equation}
We select the physical root continuously connected to the homogeneous
state at onset.
The matched homogeneous parameters are then
\begin{equation}
 \kappa_I^{\rm hom}=\frac{q_I^2}{2h_{\rm hom}^2},\qquad
 s_{\rm hom}=\left(\frac{h_{\rm hom}}
 {\sqrt{\prod_I(1+\kappa_I^{\rm hom})}}\right)^{1/3}.
 \label{eq:homreconstruction}
\end{equation}
Their temperature follows from \eqref{eq:homthermo}. Neither $s_{\rm hom}$
nor $T_{\rm hom}$ is set equal to its onset value or to the corresponding
striped parameter. We compute
\begin{equation}
 \Delta h=h-h_{\rm hom}(e,q_I),\qquad
 \frac{\Delta S}{S_{\rm hom}}=\frac{\Delta h}{h_{\rm hom}}.
 \label{eq:comparison}
\end{equation}
Energy varies along our fixed-charge continuation, but each comparison
pair has exactly the same energy.

The leading entropy difference follows from the first law. Expand
\begin{align}
 e&=e_c+e_2\eps^2+e_4\eps^4+\cdots,&
 h&=h_c+h_2\eps^2+h_4\eps^4+\cdots,\\
 T&=T_c+T_2\eps^2+\cdots,&
 e_2&=4\pi T_ch_2,\quad e_4=4\pi T_ch_4+2\pi T_2h_2.
 \label{eq:expansions}
\end{align}
At fixed charges define
$T_e=(\partial T_{\rm hom}/\partial e)_{q_I}$ at onset. Matching shifts
the homogeneous temperature to
$T_{\rm hom}=T_c+T_ee_2\eps^2+O(\eps^4)$, and its entropy is
\begin{equation}
 h_{\rm hom}=h_c+\frac{e_2}{4\pi T_c}\eps^2
 +\left(\frac{e_4}{4\pi T_c}
       -\frac{T_ee_2^2}{8\pi T_c^2}\right)\eps^4+O(\eps^6).
\end{equation}
The quadratic term cancels in \eqref{eq:comparison}, leaving
\begin{equation}
 \Delta h=C_4\eps^4+O(\eps^6),\qquad
 C_4=-\frac{e_2(T_2-T_ee_2)}{8\pi T_c^2}.
 \label{eq:quartic}
\end{equation}
Symmetry explains the even powers; the first law explains the absence of
an entropy difference at second order. Equation~\eqref{eq:quartic} also
allows an extraction of the leading difference from quadratic shifts,
independently of subtracting the two nearby entropies.

\subsection{Numerical method}
\label{sec:method}
We follow the Einstein--DeTurck framework, for which
\cite{DSW} is an excellent pedagogical reference. For a reference
metric $\bar g$, define
\begin{equation}
 \xi^M=g^{PQ}(\Gamma^M_{PQ}-\bar\Gamma^M_{PQ}),\qquad
 \cE^D_{MN}=\cE_{MN}-\nabla_{(M}\xi_{N)}=0,
 \label{eq:deturck}
\end{equation}
where $\cE_{MN}=0$ denotes the trace-reversed Einstein equation including
matter and parentheses include $1/2$. For vacuum AdS$_5$ this would read
$R_{MN}+4g_{MN}-\nabla_{(M}\xi_{N)}=0$. The DeTurck term removes the
coordinate degeneracy of the highest metric derivatives. In our static
exterior their principal part is
$-\frac12\gamma^{ij}\partial_i\partial_jg_{MN}$, with $\gamma$ the positive
metric on the $(r,z)$ base. The system is therefore elliptic there.

Our reference metric is explicitly \eqref{eq:metric} with
$U_t=U_r=U_p=U_z=W=0$, using \eqref{eq:reference}. It is a homogeneous
STU black brane, not pure AdS in the interior. We solve for ten functions
and the three global parameters $s,\kappa_1,\kappa_3$.
The integrated-charge constraints are
\begin{equation}
 q_1=q_2=q_{1,c},\qquad q_3=q_{3,c},
\end{equation}
where $q_{I,c}$ are the onset values. The exchange symmetry makes the
first two charge conditions identical, leaving two independent constraints.
Together with the prescribed amplitude \eqref{eq:amplitude}, these determine
$s,\kappa_1,\kappa_3$. We prescribe a family of homogeneous reference metrics
parametrized by these three numbers. Its connection is recomputed when
they change; at fixed parameters it is independent of the ten unknown
deformation functions.

The radial representation must accommodate the gauge's boundary falloffs.
Let $v^M$ generate an infinitesimal coordinate transformation. Using the
common leading AdS metric, preservation of the linearized DeTurck condition
requires $\nabla_{\rm AdS}^2v^M+R_{\rm AdS}{}^M{}_Nv^N=0$.
Substituting $v^r\propto r^p$ into the leading near-boundary equation gives
$p^2-4p+1=0$, the indicial equation that determines the allowed powers.
The admissible root is $p=2+\sqrt3$; the other root produces a deformation
that grows at the boundary. Since the relative metric deformation behaves
as $v^r/r$, the admissible root gives $U\propto r^{1+\sqrt3}$.
Removing the $r^2$ factor in \eqref{eq:auxiliaries} leaves
$r^{\sqrt3-1}$ in the metric auxiliary functions. These fractional terms
are included in the radial basis. Fourier projection implements
\eqref{eq:fourier}, and damped Newton
iteration solves the discrete equations, starting from the linear seed.
Details are given in appendix~\ref{app:numerics}.

In static vacuum gravity with a negative cosmological constant, the
Einstein--DeTurck equations imply a differential inequality for the
nonnegative norm $\xi^2=g_{MN}\xi^M\xi^N$. Under suitable regularity and
boundary conditions, a maximum-principle argument forces $\xi^2=0$,
excluding Ricci solitons and ensuring that the solution satisfies the
original Einstein equations~\cite{DSW}. Here Ricci solitons denote solutions
of the modified equations with a nonvanishing DeTurck term.
The matter fields in our model modify this identity, and we have not
established the corresponding sign conditions. We therefore check the
DeTurck vector and the residuals of the original Einstein and matter
equations independently. A small vector alone does not control its covariant
derivatives in \eqref{eq:deturck}. Radial and Fourier refinement, electric
flux conservation, the averaged trace Ward identity and the first law
provide additional checks. The direct and first-law entropy extractions
test the small thermodynamic signal separately from the bulk residuals.

\section{Results}
\label{sec:results}
We now present the nonlinear striped solutions and their thermodynamics.
We first specify the homogeneous background and the linearized mode at the
onset of the striped branch. We then follow this branch to finite amplitude
and compare its entropy with that of the homogeneous plasma at the same
energy and charges. Finally, we examine the spatial charge profiles,
introduce a boundary order parameter, and describe the evolution of the
temperature and modulation wavelength in thermal units.

\subsection{Homogeneous background and linear seed}
\label{sec:resultsseed}
\begin{figure}[htbp]
 \centering\includegraphics[width=.68\textwidth]{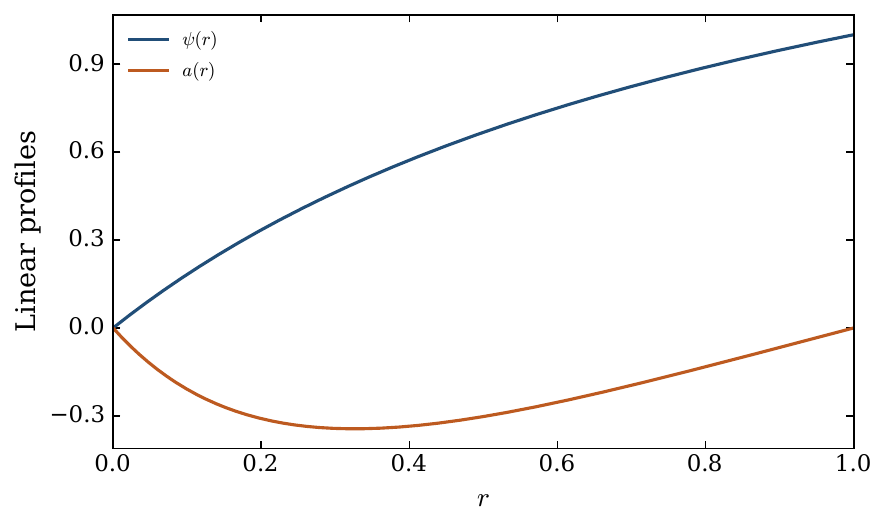}
 \caption{The static linear seed at \eqref{eq:chosenpoint}.
 The scalar and electric profiles are defined in \eqref{eq:linearansatz},
 with $\psi(1)=1$. The boundary is $r=0$ and the horizon is $r=1$.}
 \label{fig:linear}
\end{figure}

We choose the homogeneous background at onset to have
\begin{equation}
 \kappa_{1,c}=\kappa_{2,c}=1.1,\qquad \kappa_{3,c}=0.8,
 \label{eq:chosenpoint}
\end{equation}
and set $s_c=1$ to fix the overall scale. At this point the thermodynamic
Hessian has exactly one negative eigenvalue, in the exchange-odd direction
\eqref{eq:oddhessian}. The linear matching problem gives
\begin{equation}
 \frac{k_c}{2\pi T_c}\simeq0.453(1),
 \label{eq:criticalnumber}
\end{equation}
where parentheses enclose the final quoted digit, not an uncertainty
estimate. The radial profiles are shown in figure~\ref{fig:linear}. The scalar
profile has no interior nodes. We fix $L_z=2\pi/k_c$ and continue the
nonlinear branch at the onset total charges. The parameters $s,\kappa_1,\kappa_3$
then adjust with the modulation; \eqref{eq:chosenpoint} specifies the
onset rather than the whole branch.

\subsection{Entropy and the onset of the striped branch}
\label{sec:resultsentropy}
We construct the nonlinear striped phase from the linearized seed of
section~\ref{sec:linear} by progressively increasing the amplitude $\eps$,
defined in \eqref{eq:amplitude} and appearing in the linearized fields
\eqref{eq:linearansatz}.\footnote{This mirrors the construction of the
chirally broken phase of cascading gauge theory in de Sitter,
where a linearized symmetry-breaking mode seeds the nonlinear
branch~\cite{CascadingDeSitter}.}
At each amplitude we solve the full nonlinear equations, keeping the torus
periods and total charges fixed. The energy changes along the resulting
branch, and we compare each striped solution with the homogeneous state
at the same energy and charges.

\begin{figure}[htbp]
 \centering\includegraphics[width=\textwidth]{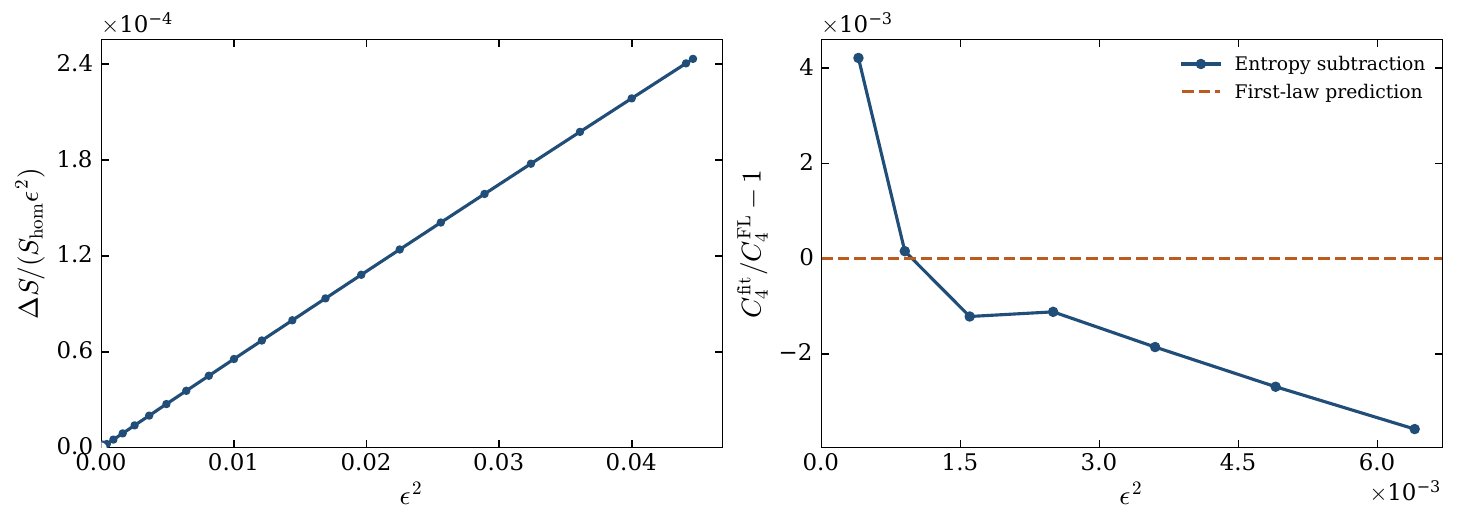}
 \caption{Left: $\Delta S/(S_{\rm hom}\eps^2)$ versus $\eps^2$, with
 the entropies compared at matched $E,Q_I,V$. The open point denotes the
 zero-amplitude limit. The quartic scaling in \eqref{eq:quartic} gives
 a straight line near the origin with slope $C_4/h_c$.
 Right: $C_4^{\rm fit}/C_4^{\rm FL}-1$ versus $\eps^2$ near onset,
 where $C_4^{\rm fit}(\eps)\equiv\Delta h(\eps)/\eps^4$ is the
 finite-amplitude estimate and $C_4^{\rm FL}$ is the first-law coefficient
 in \eqref{eq:quartic}. The dashed line denotes agreement with that
 prediction; higher-order corrections begin at $O(\eps^2)$.
 Symbols are numerical solutions; connecting lines guide the eye.}
 \label{fig:entropy}
\end{figure}

The striped branch extends towards lower energy. Its entropy excess is positive throughout the sampled
branch, as shown in figure~\ref{fig:entropy}. The striped phase is therefore
entropically favored over its homogeneous microcanonical comparator.
The difference vanishes continuously at onset.

The left panel displays the quartic behavior derived in
\eqref{eq:quartic} as linear growth near the origin. The right panel
shows the relative deviation of the finite-amplitude coefficient from
the first-law prediction. Extrapolating the direct entropy subtraction to $\eps=0$ gives a positive
$C_4$ that agrees with the first-law extraction (see eq.~\eqref{eq:quartic})
to approximately $1\%$. Thus the entropy preference already appears on the
weakly modulated branch.

The sign of the leading entropy difference can be understood from the two
factors in \eqref{eq:quartic}. Along the striped branch,
\begin{equation}
 e-e_c=e_2\eps^2+O(\eps^4),\qquad e_2<0,
\end{equation}
so its energy decreases as the amplitude grows away from onset. At each
amplitude, the homogeneous comparison state has the same energy and charges
as the striped state. Its temperature therefore shifts according to
\begin{equation}
 T_{\rm hom}=T_c+T_ee_2\eps^2+O(\eps^4),\qquad
 T_e=\left.\left(\frac{\partial T_{\rm hom}}{\partial e}\right)_{q_I}\right|_{e=e_c}.
\end{equation}
The temperature difference between the matched states is thus
\begin{equation}
 T-T_{\rm hom}=(T_2-T_ee_2)\eps^2+O(\eps^4).
\end{equation}
Equation~\eqref{eq:quartic},
\begin{equation}
 C_4=-\frac{e_2(T_2-T_ee_2)}{8\pi T_c^2},
\end{equation}
then shows that, because $e_2<0$, the positive leading entropy difference
corresponds to the striped state having a higher temperature than its
homogeneous counterpart. The decrease in energy refers to motion along
the branch from onset; the entropy comparison always involves states at
identical energy and charges.

\subsection{Charge profiles and a boundary order parameter}
\label{sec:resultsprofiles}
We next characterize the spatial structure of the striped phase through
the local charge densities. Their profiles show how the conserved charges
redistribute along the circle as the modulation grows. We also introduce
a dimensionless boundary order parameter that measures the charge-density
contrast and tracks the continuous departure from the homogeneous state.

\begin{figure}[htbp]
 \centering\includegraphics[width=\textwidth]{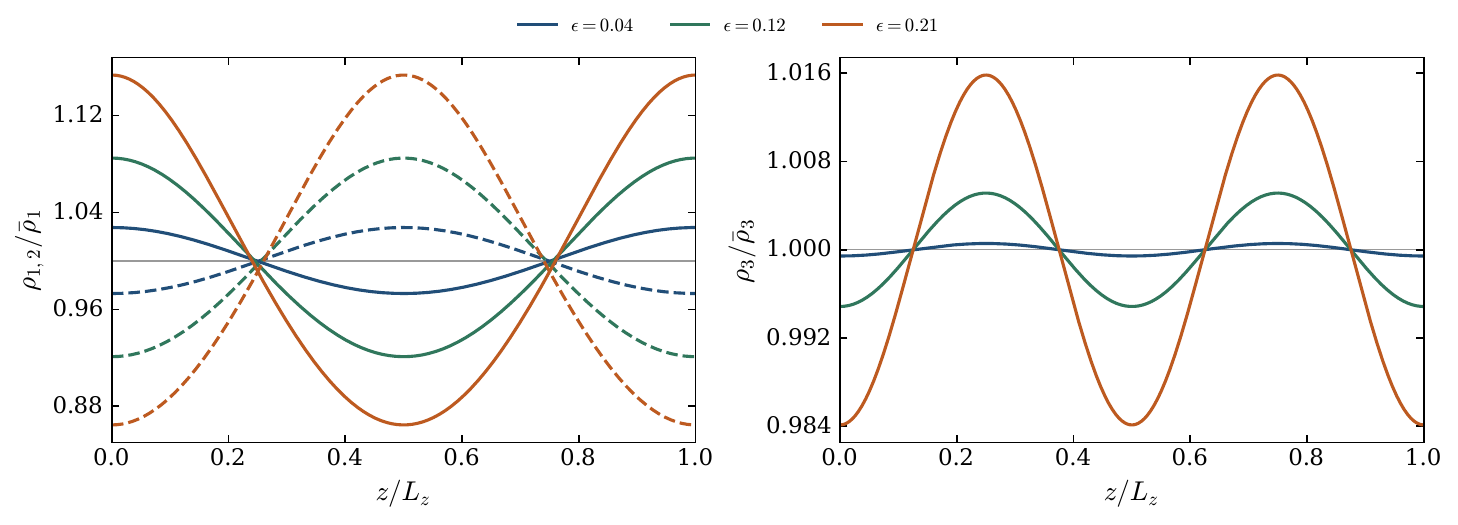}
 \caption{Boundary charge profiles for $\eps=0.04,0.12,0.21$.
 Left: $\rho_1/\bar\rho_1$ (solid) and $\rho_2/\bar\rho_1$ (dashed),
 with $\bar\rho_1=\bar\rho_2$. Right: $\rho_3/\bar\rho_3$.
 The coordinate $z/L_z$ spans one full stripe period.}
 \label{fig:profiles}
\end{figure}

Figure~\ref{fig:profiles} shows three representative solutions, normalized
by their fixed spatial averages $\bar\rho_I=Q_I/V$.
Charges 1 and 2 accumulate in opposite parts of the circle. Their profiles
are related by a half-period translation, while the third charge has
period $L_z/2$. Its modulation is the even response to the primary
charge-difference mode. As the amplitude grows, the profiles depart
smoothly from the linear sinusoid through the additional harmonics allowed
by \eqref{eq:fourier}.

\begin{figure}[htbp]
 \centering\includegraphics[width=\textwidth]{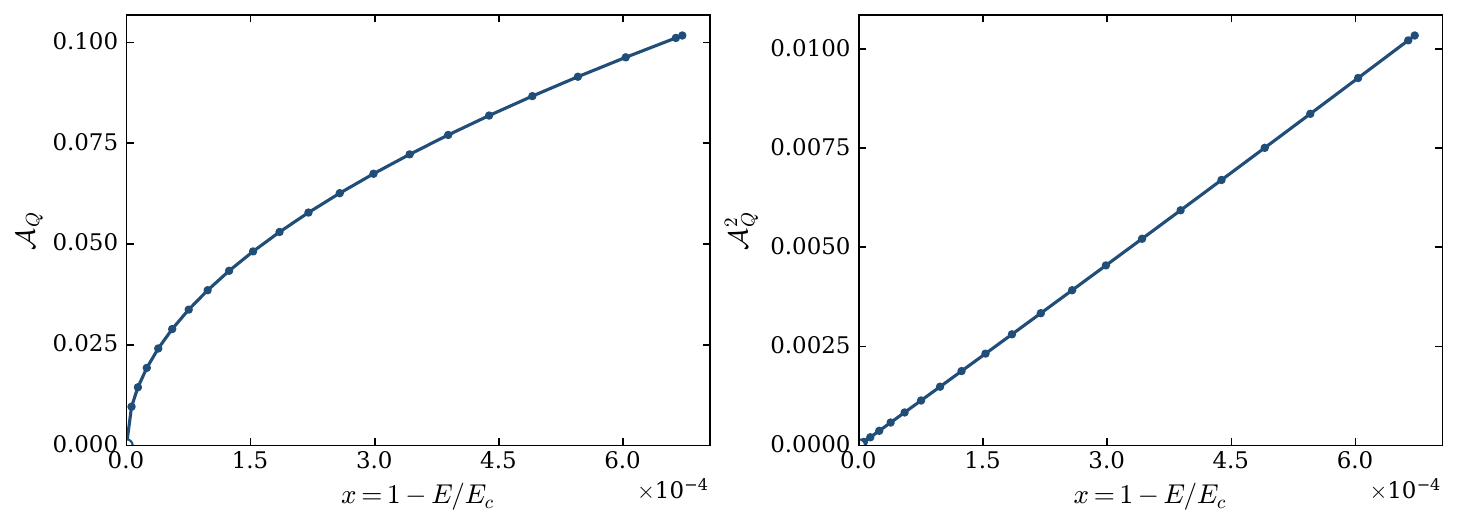}
 \caption{The boundary charge contrast \eqref{eq:contrast} and its square
 as functions of the dimensionless energy displacement $x=1-E/E_c$.
 Open points
 denote the homogeneous onset.}
 \label{fig:order}
\end{figure}

To characterize the stripe by a boundary observable, define its
root-mean-square charge contrast,
\begin{equation}
 \mathcal A_Q=\frac1{\bar\rho_1}
 \sqrt{\avg{\left(\frac{\rho_1-\rho_2}{2}\right)^2}}.
 \label{eq:contrast}
\end{equation}
It is dimensionless and independent of the choice of stripe origin.
The homogeneous state has $\mathcal A_Q=0$. To display its growth along
the branch, we use the dimensionless energy displacement
\begin{equation}
 x=1-\frac{E}{E_c}=1-\frac{e}{e_c},
 \label{eq:energyaxis}
\end{equation}
where $E_c$ is the onset energy at the same charges and volume.
The displayed solutions have $x>0$. Figure~\ref{fig:order} shows the
charge contrast as a function of $x$.

Near onset, the charge difference is linear in the amplitude $\eps$, while
the energy shift is quadratic:
\begin{equation}
 \mathcal A_Q=a_1|\eps|+O(|\eps|^3),\qquad
 x=-\frac{e_2}{e_c}\eps^2+O(\eps^4),
\end{equation}
where $a_1>0$ is the leading coefficient of the charge contrast and $e_2<0$
describes the decrease in energy along the striped branch. These powers
follow from the symmetry exchanging charges 1 and 2: the charge difference
changes sign under $\eps\to-\eps$, whereas the energy is invariant. The
root-mean-square contrast measures the magnitude of this charge difference.
Eliminating $\eps$ gives
\begin{equation}
 \mathcal A_Q=a_1\sqrt{\frac{e_c}{-e_2}}\,x^{1/2}+O(x^{3/2}).
\end{equation}
Thus the order parameter grows continuously with the mean-field critical
exponent $\beta=\tfrac12$, defined by $\mathcal A_Q\propto x^\beta$. Here the
control parameter is the reduced energy displacement $x$, with the charges
and spatial volume held fixed. Equivalently, $\mathcal A_Q^2$ grows linearly
with $x$ near onset, as illustrated in figure~\ref{fig:order}.

\subsection{Temperature and wavelength}
\label{sec:resultsthermal}
The temperature changes along the striped branch even though the spatial
period is held fixed. Consequently, the modulation wavelength measured in
thermal units also changes. We examine this variation and compare the
striped temperature with that of the homogeneous state at the same energy
and charges, relating their difference to the growth of the entropy
advantage.

\begin{figure}[htbp]
 \centering\includegraphics[width=\textwidth]{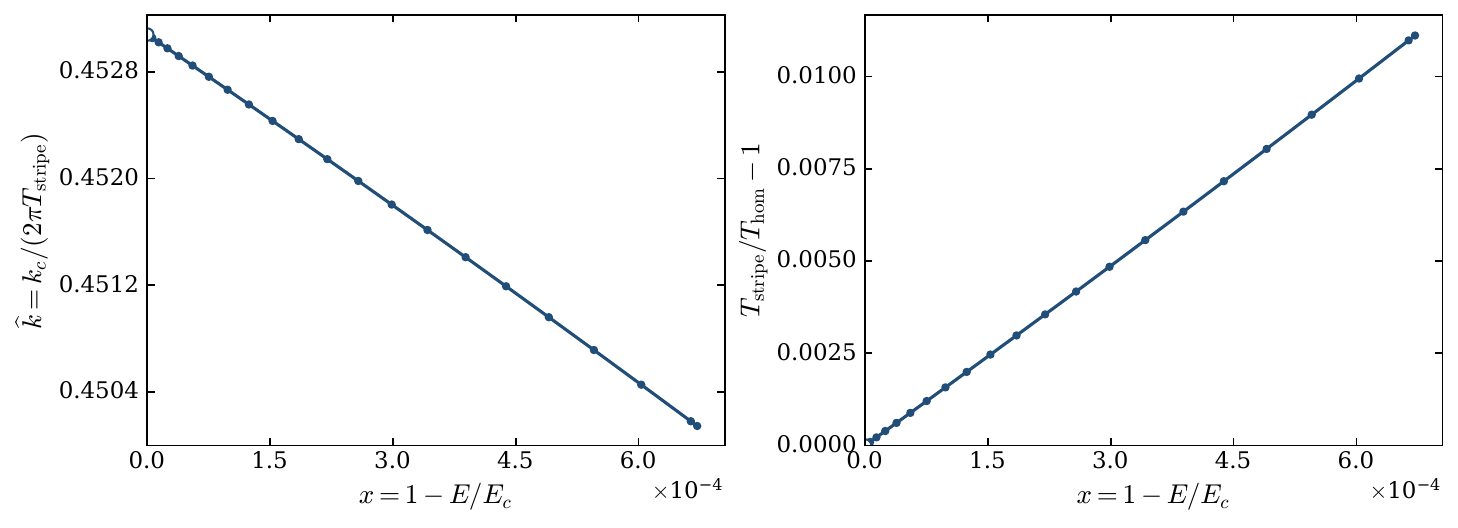}
 \caption{Left: stripe momentum in units of its own temperature.
 Right: the stripe temperature relative to that of the homogeneous state
 with the same energy and charges. Both panels use $x=1-E/E_c$ as the
 horizontal variable. The temperatures coincide at the
 onset and differ along the nonlinear branch.}
 \label{fig:thermal}
\end{figure}

Figure~\ref{fig:thermal} shows the wave number in thermal units,
\begin{equation}
 \widehat k=\frac{k_c}{2\pi T_{\rm stripe}}.
\end{equation}
The stripe temperature increases along the sampled branch, so
$\widehat k$ decreases. The corresponding thermal circumference is
$\widehat L_z=2\pi/\widehat k$.

The matched homogeneous temperature instead decreases as the energy
decreases along this segment. The stripe is consequently hotter than its comparator,
as shown in the right panel. This is consistent with the entropy plot:
at fixed charges,
\begin{equation}
 \frac{\dd\Delta h}{\dd e}
 =\frac1{4\pi}\left(\frac1{T_{\rm stripe}}-
                         \frac1{T_{\rm hom}}\right)<0.
\end{equation}
Starting from zero at $e=e_c$, the entropy excess therefore grows as
the energy decreases along the striped branch.

\FloatBarrier

\section{Discussion}
\label{sec:discussion}
We have constructed nonlinear striped states of charged
$\mathcal N=4$ SYM that connect continuously to a static GIKS threshold and
have higher entropy than the homogeneous state at matched energy and
charges. The torus and the unequal charge parameters isolate a single
internal unstable direction and a simple spatial modulation. This gives
a concrete equilibrium candidate for the redistribution of charge caused
by the instability. It also provides initial and final-state benchmarks
for a real-time microcanonical evolution.

The compactification is a useful way to organize the unstable modes.
For a specified homogeneous state, define
\begin{equation}
 \widehat k=\frac{k}{2\pi T},\qquad
 \widehat L_i=2\pi TL_i,\qquad
 \widehat k_n=\frac{2\pi n}{\widehat L_z}.
 \label{eq:hatted}
\end{equation}
If its unstable band is $0<\widehat k<\widehat k_c$, no nonzero
longitudinal mode is in that band for
$\widehat L_z<2\pi/\widehat k_c$. This is a linear statement about that
homogeneous state. The cutoff also depends on its charge parameters.
On the nonlinear branch the coordinate wave number stays fixed, while
$\widehat k$ changes with the stripe temperature, as shown in
figure~\ref{fig:thermal}. Comparing different states requires stating
which temperature defines the hatted quantities.

Although constructed on a three-torus, our striped solutions also define
equilibrium states in the decompactification limit. Let $\ell$ denote the
spatial period of a given solution. We may place $n$ identical copies on
a circle of length $L_z=n\ell$, leaving the local bulk geometry and matter
fields unchanged. Taking $n\to\infty$ at fixed $\ell$ produces a periodic
striped state on the noncompact $z$-axis. The transverse directions can
likewise be decompactified because the solution is independent of $x,y$.
In this limit the total energy, charges and entropy are extensive, while
their spatially averaged densities remain fixed. Consequently, the entropy
advantage over the homogeneous state at the same average energy and charge
densities survives decompactification. Our results therefore establish
that, for the states studied here, the homogeneous plasma in flat space is
not the most entropic equilibrium phase. They do not exclude other
equilibria with greater entropy and stronger spatial clumping, including
configurations with longer modulation periods or dependence on the
transverse coordinates. Identifying such phases and determining the
stability of the periodic stripes are interesting questions for future work.

Repeating a fixed-period solution is different from increasing the period
of a single stripe without bound. The larger domain also admits
perturbations that distinguish neighboring cells, which are excluded by
a one-cell calculation.

A simple example is a perturbation with twice the original period:
\begin{equation}
 \chi_0(r,z)\simeq\eps\psi(r)\cos(k_sz),\qquad
 \delta\chi(r,z)=\eta v(r)\cos(k_sz/2),\qquad k_s=2\pi/\ell.
 \label{eq:doubling}
\end{equation}
Here $\eta$ is an infinitesimal perturbation amplitude. At cell centers
$z=j\ell$, the new contribution alternates in sign, allowing successive
cells to differ. The coupled perturbation includes the other bulk fields,
and nonlinear terms generate further harmonics. All momenta on a finite
circle remain commensurate; changing the common temperature normalization
does not change their ratios.

Near onset, a weak stripe repeated twice admits momentum $k_c/2$, which
lies inside the homogeneous unstable band. Continuity of a strictly
positive growth rate implies that this instability persists for
sufficiently weak repeated stripes. This does not locate a secondary
threshold along a finite-amplitude branch. In particular, the second
harmonic entering the \emph{homogeneous} unstable band at $2L_{z,c}$ is not
a general criterion for a more complicated nonlinear modulation.

The appropriate future calculation is a perturbation analysis of the
striped background. A useful form is
\begin{equation}
 \delta F(t,r,z)=e^{\sigma t+iqz}
    \sum_{m\in\mathbb Z}f_m(r)e^{imk_sz},
 \label{eq:bloch}
\end{equation}
where $F$ denotes a bulk field, $q$ is the Bloch momentum defined modulo
$k_s$, and $\operatorname{Re}\sigma>0$ signals growth. On $n$ cells, $q$ is
discrete; it becomes continuous in the many-cell limit. Period doubling
corresponds to $q=k_s/2$. This analysis can identify changes of stripe
amplitude from cell to cell, long-wavelength displacement modes, and
secondary stationary branches. Combined with construction and entropy comparison of the resulting
branches, this analysis would help identify preferred inhomogeneous states
at the same conserved quantities.

Increasing the transverse periods opens further channels. Perturbations
can bend the stripes or modulate their amplitude in $x,y$, and their
nonlinear development could produce two- or three-dimensional structures.
Domain patterns in magnetic films and modulated phases of block copolymers
provide familiar morphological examples~\cite{Saratz,MatsenSchick}.
Here the charge-transport dynamics and gravitational equations must decide
which patterns occur. Their stability, construction and competition in
the decompactification limit are natural extensions of the present work.

\acknowledgments
\begingroup\interlinepenalty=10000
I would like to thank Ofer Aharony for discussions many years ago regarding
striped phases in holography. This work was supported by the Natural Sciences
and Engineering Research Council of Canada (NSERC).
\par\endgroup

\appendix
\section{STU action and homogeneous solution}
\label{app:stu}
We set the AdS radius to one. In the conventions used here, the STU
action in the sector of interest is
\begin{equation}
 I=\frac1{16\pi G_5}\int\dd^5x\sqrt{-g}\left[
 R-\tfrac12(\partial\phi)^2-\tfrac12(\partial\chi)^2
 +4\sum_I(X^I)^{-1}-\tfrac18\sum_I(X^I)^{-2}(F^I)^2\right]+I_{\rm CS},
 \label{eq:action}
\end{equation}
where $F^I=\dd A^I$ and
\begin{equation}
 X^1=e^{-\phi/\sqrt6-\chi/\sqrt2},\qquad
 X^2=e^{-\phi/\sqrt6+\chi/\sqrt2},\qquad X^3=e^{2\phi/\sqrt6},\qquad
 X^1X^2X^3=1.
\end{equation}
The Chern--Simons term $I_{\rm CS}$ and its field-equation contributions
vanish for \eqref{eq:electric}. Let
$V_+=4\sum_I(X^I)^{-1}$ and $Z_I=(X^I)^{-2}$. The equations retained are
\begin{align}
 \cE_{MN}={}&R_{MN}
 -\tfrac12\partial_M\phi\partial_N\phi
 -\tfrac12\partial_M\chi\partial_N\chi+\tfrac13V_+g_{MN}\\
 &-\tfrac14\sum_I Z_I
       (F^I_{MP}F_N^{I\ P}-\tfrac16g_{MN}(F^I)^2)=0,\nonumber\\
 0={}&\nabla^2\varphi+\partial_\varphi V_+
       -\tfrac18\sum_I(\partial_\varphi Z_I)(F^I)^2,
       \qquad\varphi=\phi,\chi,\\
 0={}&\nabla_M(Z_IF^{I\,MN}).
 \label{eq:eoms}
\end{align}

For general homogeneous parameters, put
\begin{equation}
 H_I=1+\kappa_Ir,\qquad H=H_1H_2H_3,\qquad
 P=\prod_I(1+\kappa_I),\qquad f=H-Pr^2.
\end{equation}
The solution is
\begin{align}
 \dd s^2&=-\frac{s^2f}{rH^{2/3}}\dd t^2
       +\frac{s^2H^{1/3}}r(\dd x^2+\dd y^2+\dd z^2)
       +\frac{H^{1/3}}{4r^2f}\dd r^2,\\
 X^I&=\frac{H^{1/3}}{H_I},\qquad
 \bar\Phi_I=s\sqrt{2\kappa_IP}
       \left[\frac1{1+\kappa_I}-\frac r{1+\kappa_Ir}\right].
 \label{eq:stusolution}
\end{align}
In particular, when $\kappa_1=\kappa_2$,
\begin{equation}
 \bar\phi=\frac{\sqrt6}{2}\left(\tfrac13\log H-\log H_3\right),
 \qquad\bar\chi=0.
\end{equation}
Horizon regularity, electric flux and the
renormalized boundary stress tensor give \eqref{eq:homthermo}.
For the standard AdS$_5\times S^5$ normalization $G_5=\pi/(2N^2)$,
\begin{equation}
 \frac EV=\frac{N^2}{8\pi^2}e,\qquad
 \frac{Q_I}V=\frac{N^2}{8\pi^2}q_I,\qquad
 \frac SV=\frac{N^2}{2\pi}h.
\end{equation}

\section{Numerical implementation and checks}
\label{app:numerics}
\subsection{Radial representation and Fourier projection}
The radial nodes are interior Chebyshev--Gauss points,
\begin{equation}
 r_i=\frac{1-\cos[(i+\tfrac12)\pi/N_r]}2,
 \qquad i=0,\ldots,N_r-1.
\end{equation}
For each Fourier harmonic, an auxiliary field is expanded in the
$N_r$-dimensional space
\begin{equation}
 u(r)=\sum_{j=0}^{N_r-4}c_jT_j(2r-1)
       +\sum_{j=0}^2d_jr^{\beta+j}.
 \label{eq:basis}
\end{equation}
Here $T_j$ are Chebyshev polynomials, $\beta=\sqrt3-1$ for metric auxiliaries,
and $\beta=\sqrt3+1$ for scalar and electric auxiliaries. The latter powers
follow by applying the same radial coordinate deformation to the leading
matter fields. All these basis functions are analytic near the horizon.
The factors in \eqref{eq:auxiliaries} impose the leading boundary and
conical conditions; the regular function space excludes the singular
horizon branches. The equations at the interior nodes determine the
remaining endpoint relations. No finite radial cutoff is introduced.

If $b_j(r)$ lists the basis functions and $V_{ij}=b_j(r_i)$, the nodal
derivative matrices are
\begin{equation}
 D=V'V^{-1},\qquad D^{(2)}=V''V^{-1}.
\end{equation}
Boundary and horizon values are evaluated from the same radial interpolant.
The second derivative is constructed directly because differentiation of
a fractional power need not stay within the interpolation space.
Matrix construction uses increased precision to control the near linear
dependence of the polynomial and fractional-power basis functions.

We truncate \eqref{eq:fourier} at a maximum harmonic $m_{\max}$.
Nonlinear residuals are evaluated on an oversampled uniform angular grid
and projected onto the retained cosine or sine modes. For example,
\begin{equation}
 R_0(r)=\frac1{N_z}\sum_jR(r,\theta_j),\qquad
 R_m(r)=\frac2{N_z}\sum_jR(r,\theta_j)\cos(m\theta_j),\quad m>0.
\end{equation}
The sine projection is used for $W$. Derivatives of the retained Fourier
modes are exact; the oversampled grid controls aliasing from nonlinear
products. Independent Fourier refinement tests the omitted harmonics.
The displayed branch uses $(N_r,m_{\max})=(46,11)$, with the
$\eps=0.06$ point taken from $(54,13)$. Independent refinements use
$(54,11)$, $(46,13)$, and $(54,13)$, with $N_z=4m_{\max}+8$.

\subsection{Equations and nonlinear iteration}
Write the static metric as
\begin{equation}
 \dd s^2=-e^{2a}\dd t^2+e^{2b}(\dd x^2+\dd y^2)
          +\gamma_{ij}\dd x^i\dd x^j,\qquad (x^i)=(r,z).
\end{equation}
For \eqref{eq:metric}, $a=\log\bar N+U_t$ and
$b=\frac12\log\bar G_s+U_p$. If $c=\frac12\log\bar G_s+U_z$ and
$d=\frac12\log\bar G_r+U_r$, then
\begin{equation}
 \gamma^{rr}=e^{-2d},\qquad\gamma^{rz}=-We^{-2d},\qquad
 \gamma^{zz}=e^{-2c}+W^2e^{-2d}.
\end{equation}
Writing $\Delta_\gamma$ for the scalar Laplacian of the two-dimensional
metric $\gamma_{ij}$, the scalar Laplacian and electrostatic Maxwell
equation reduce to
\begin{align}
 \nabla^2\varphi&=\Delta_\gamma\varphi+
      \gamma^{ij}\partial_i\varphi\partial_j(a+2b),\\
 0&=\Delta_\gamma\Phi_I+
      \gamma^{ij}\partial_i\Phi_I\partial_j(-a+2b+\log Z_I).
\end{align}
These are combined with the five independent Einstein--DeTurck components.
The DeTurck vector includes the contributions from the warped $t,x,y$
directions, not only the connection of the two-dimensional base.
Known vanishing endpoint factors are divided out of the residuals before
projection, so they do not mask errors near the boundary or horizon.

For the numerical iteration, we parametrize the three global unknowns by
\begin{equation}
 \lambda_s=\log(s/s_c),\qquad
 \lambda_1=\log(\kappa_1/\kappa_{1,c}),\qquad
 \lambda_3=\log(\kappa_3/\kappa_{3,c}).
 \label{eq:globalunknowns}
\end{equation}
Let $\mathbf u$ contain the retained field values and these three
logarithmic parameters. The residual vector $\mathbf R(\mathbf u)$
contains the projected field equations, two fixed-charge conditions and
$C_1(1)-\eps=0$. At each Newton step we solve
\begin{equation}
 J\delta\mathbf u=-\mathbf R,\qquad
 J_{ab}=\frac{\partial R_a}{\partial u_b},\qquad
 \mathbf u_{\rm new}=\mathbf u+\alpha\delta\mathbf u.
\end{equation}
The step length $\alpha$ is reduced until the nonlinear residual decreases.
The first solution is initialized with the linear seed, and subsequent
solutions use continuation in $\eps$. The amplitude condition prevents
convergence back to the homogeneous solution.

\subsection{Validation}
\label{app:validation}
The homogeneous limit reproduces the analytic STU solution and its
thermodynamics. We check the nonlinear solutions under radial and Fourier
refinement, and reevaluate the unprojected equations on a denser grid.
The latter is an interpolation audit, distinct from solving again at a
higher resolution. The original Einstein equations, scalar and Maxwell
equations, DeTurck vector and radial electric-flux conservation are
evaluated separately.

The implemented stress-tensor check is the spatially averaged trace Ward
identity. The energy obtained from the boundary stress tensor also agrees
with its variation inferred from the fixed-charge first law. This is an
independent test because the first law is not used to define the energy.
Near onset, fits over several amplitude windows give the same sign of
$C_4$ from entropy subtraction and from \eqref{eq:quartic}. The direct
subtraction is more sensitive to boundary coefficient extraction; the
spread of its extrapolated coefficient is at the percent level rather
than the apparent precision of the horizon data.

\begin{figure}[htbp]
 \centering\includegraphics[width=.69\textwidth]{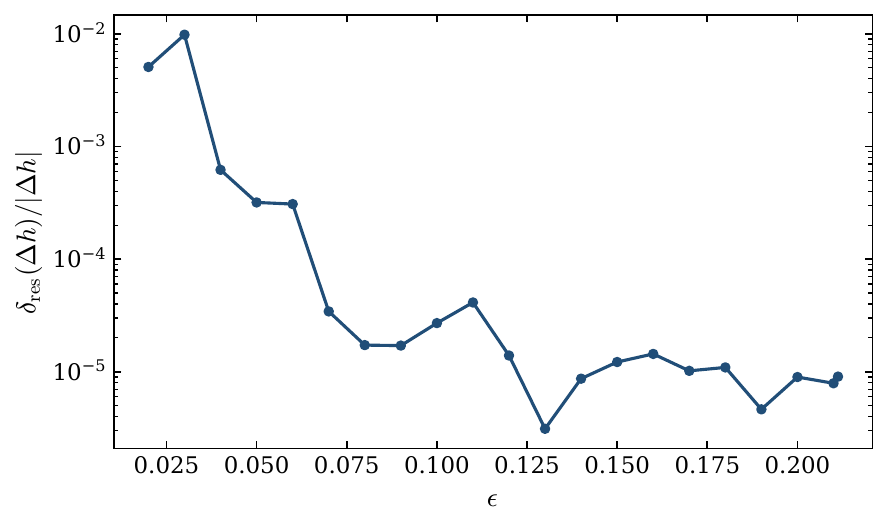}
 \caption{Relative change of the matched entropy difference under resolution
 refinement. Here $\Delta h$ is defined in \eqref{eq:comparison}.
 The quantity $\delta_{\rm res}(\Delta h)/|\Delta h|$ is the largest
 available paired-grid difference divided by the magnitude of $\Delta h$
 on the finer grid of that pair. Independently relaxed refinement pairs
 are used where available, including comparisons that did not meet the
 tighter refinement target; elsewhere the original resolution pair is used.}
 \label{fig:convergence}
\end{figure}

Figure~\ref{fig:convergence} compares the matched entropy difference across
paired resolutions. Where independently relaxed radial and Fourier
refinements are available, we display the largest relative difference.
The observed spread reaches about one percent at small amplitude and
remains small relative to the positive entropy signal. Together
with the independent thermodynamic checks they support the entropic
preference reported here. They are deterministic convergence diagnostics,
not statistical error bars.
\FloatBarrier

\bibliographystyle{JHEP}
\bibliography{references}
\end{document}